\documentclass[aps,twocolumn,superscriptaddress,floatfix,prb]{revtex4-1}

\usepackage[utf8]{inputenc}
\usepackage{amsmath}
\usepackage{amssymb}
\usepackage[colorlinks=true,citecolor=blue,linkcolor=blue,urlcolor=blue]{hyperref}
\usepackage[capitalise]{cleveref}

\usepackage{times}
\usepackage[english]{babel}

\usepackage{braket}

\usepackage{graphicx}
\usepackage[export]{adjustbox}

\usepackage{xcolor}
\newcommand{\dashSec}[1]{\textit{#1}.---}
\newcommand{\eifrac}[2]{#1/\left(#2 \kern 0.05em \right)}

\begin{document}

\title{Quantum network transfer and storage with compact localized states induced by local symmetries}

\author{M. Röntgen}
\affiliation{%
	Zentrum für optische Quantentechnologien, Universität Hamburg, Luruper Chaussee 149, 22761 Hamburg, Germany
}%

\author{C. V. Morfonios}%
\affiliation{%
	Zentrum für optische Quantentechnologien, Universität Hamburg, Luruper Chaussee 149, 22761 Hamburg, Germany
}%

\author{I. Brouzos}%
\affiliation{%
	Department of Physics, University of Athens, 15771 Athens, Greece
}%

\author{F. K. Diakonos}%
\affiliation{%
	Department of Physics, University of Athens, 15771 Athens, Greece
}%

\author{P. Schmelcher}
\affiliation{%
	Zentrum für optische Quantentechnologien, Universität Hamburg, Luruper Chaussee 149, 22761 Hamburg, Germany
}%
\affiliation{%
	The Hamburg Centre for Ultrafast Imaging, Universität Hamburg, Luruper Chaussee 149, 22761 Hamburg, Germany
}%

\begin{abstract}
We propose modulation protocols designed to generate, store and transfer compact localized states in a quantum network.
Induced by parameter tuning or local reflection symmetries, such states vanish outside selected domains of the complete system and are therefore ideal for information storage.
Their creation and transfer is here achieved either via amplitude phase flips or via optimal temporal control of inter-site couplings.
We apply the concept to a decorated, locally symmetric Lieb lattice where one sublattice is dimerized, and also demonstrate it for more complex setups.
The approach allows for a flexible storage and transfer of states along independent paths in lattices supporting flat energetic bands.
The generic network and protocols proposed can be utilized in various physical setups such as atomic or molecular spin lattices, photonic waveguide arrays, and acoustic setups.
\end{abstract}

\maketitle

\dashSec{Introduction}The Storage and transfer of information in quantum systems is a task of great importance for  the realization of quantum computers and simulators.
Storage of a quantum state implies that it is effectively decoupled from surrounding building blocks of a system, thus not affected by its environment.
In contrast, transfer of a state requires a successive interaction with its environment leading to its directed propagation.
We here propose a quantum state that can be easily prepared and robustly stored and present protocols that transfer it through a quantum network.
We thereby merge key ingredients of three different fields of research: 
(i) compact localized states in flat band lattices \cite{Maimaiti2017PRB95115135CompactLocalizedStatesFlatband,Leykam2018APX31473052ArtificialFlatBandSystems,Mukherjee2015PRL114245504ObservationLocalizedFlatBandState,Taie2015SA1CoherentDrivingFreezingBosonic,Apaja2010PRA8241402FlatBandsDiracCones,Mukherjee2015OLO405443ObservationLocalizedFlatbandModes,Derzhko2015IJMPB291530007StronglyCorrelatedFlatbandSystems,Vicencio2015PRL114245503ObservationLocalizedStatesLieb,Morales-Inostroza2016PRA9443831SimpleMethodConstructFlatband,Rontgen2018PRB9735161CompactLocalizedStatesFlat,Wan2017SR715188HybridInterferenceInducedFlat,Shen2010PRB8141410SingleDiracConeFlat,Real2017SR715085FlatbandLightDynamicsStub,Rojas-Rojas2017PRA9643803QuantumLocalizedStatesPhotonic}, 
(ii) perfect state transfer \cite{Christandl2004PRL92187902PerfectStateTransferQuantum,Christandl2005PRA7132312PerfectTransferArbitraryStates,Bose2007CP4813QuantumCommunicationSpinChain,Kay2010IJQI08641PerfectEfficientStateTransfer,Cirac1997PRL783221QuantumStateTransferEntanglement,Yung2005PRA7132310PerfectStateTransferEffective,Wojcik2005PRA7234303UnmodulatedSpinChainsUniversal,Osborne2004PRA6952315PropagationQuantumInformationSpin,Chapman2016NC711339ExperimentalPerfectStateTransfer}, and 
(iii) optimal control theory  \cite{Brif2010NJP1275008ControlQuantumPhenomenaPast,Glaser2015EPJD69279TrainingSchrodingersCatQuantum,Murphy2010PRA8222318CommunicationQuantumSpeedLimit,Zhang2016AoP375435OptimalControlFastHighfidelity,Nakao2017JPBAMOP5065501OptimalControlPerfectState}.
%\cm{Should we really put all these gathered references here like this?}

A compact localized state (CLS) is a Hamiltonian eigenstate defined by its strictly vanishing amplitudes outside a spatial subdomain of the system.
This compact localization originates from destructive interference caused by the right combination of lattice geometry and Hamiltonian matrix elements.
Such a combination is possible in a broad range of physical systems \cite{Leykam2018APX31473052ArtificialFlatBandSystems}, and CLSs have been realized in, e.g., photonic waveguide arrays \cite{Mukherjee2015PRL114245504ObservationLocalizedFlatBandState,Vicencio2015PRL114245503ObservationLocalizedStatesLieb,Mukherjee2015OLO405443ObservationLocalizedFlatbandModes,Real2017SR715085FlatbandLightDynamicsStub}, ultracold atomic ensembles \cite{Taie2015SA1CoherentDrivingFreezingBosonic,Apaja2010PRA8241402FlatBandsDiracCones}, and optomechanical setups \cite{Wan2017SR715188HybridInterferenceInducedFlat}.
Though typically residing in dispersionless---or `flat'---energy bands of periodic lattices with macroscopic degeneracy \cite{Mukherjee2015PRL114245504ObservationLocalizedFlatBandState,Taie2015SA1CoherentDrivingFreezingBosonic,Vicencio2015PRL114245503ObservationLocalizedStatesLieb,Mukherjee2015OLO405443ObservationLocalizedFlatbandModes,Shen2010PRB8141410SingleDiracConeFlat,Apaja2010PRA8241402FlatBandsDiracCones,Real2017SR715085FlatbandLightDynamicsStub,Rojas-Rojas2017PRA9643803QuantumLocalizedStatesPhotonic}, CLSs can exist in non-periodic setups just as well.

By their defining property, CLSs are ideally suited for storage:
Due to their compactness, they can be stored using only a very small number of physical sites and, being Hamiltonian eigenstates, they can in principle be stored for an infinite amount of time.
Moreover, their compactness protects CLSs against a wide range of imperfections. In particular, CLSs remain unaffected by changes of the Hamiltonian outside their localization domain.
Furthermore, there is a class of CLSs which are protected by spatial \textit{local symmetry} of the Hamiltonian against \textit{any} perturbations preserving this local symmetry and the geometry within their localization domain \cite{Rontgen2018PRB9735161CompactLocalizedStatesFlat}.
As a side note, the more general study of local symmetries has recently been put on new grounds by introducing a framework of non-local currents \cite{Kalozoumis2013PRA8732113LocalSymmetriesOnedimensionalQuantum,Kalozoumis2015AP362684InvariantCurrentsScatteringLocally,Spourdalakis2016PRA9452122GeneralizedContinuityEquationsTwofield,Morfonios2017AP385623NonlocalDiscreteContinuityInvariant,Rontgen2017AP380135NonlocalCurrentsStructureEigenstates,Zampetakis2016JPA49195304InvariantCurrentApproachWave} by means of which the parity and Bloch-theorem are generalized to locally symmetric systems \cite{Kalozoumis2014PRL11350403InvariantsBrokenDiscreteSymmetries}.
Recently, it has been shown that CLSs may result from particular local symmetries which commute with the Hamiltonian in discrete systems \cite{Rontgen2018PRB9735161CompactLocalizedStatesFlat}.

While favoring their storage, the compactness and consequent decoupling of CLSs from their surroundings poses the challenge of how to \textit{transfer} them controllably across a lattice.
We here demonstrate how transfer of local-symmetry induced CLSs can be achieved using two different approaches, based on free and driven time-evolution.
The first approach utilizes the common perfect (i.e., with unit fidelity) quantum state transfer scenario, where static inter-site couplings are tailored such that a selected state evolves freely from one location to another.
Quantum state transfer techniques are especially explored in connection with entanglement transfer \cite{Kay2010IJQI08641PerfectEfficientStateTransfer,Bose2007CP4813QuantumCommunicationSpinChain}, while engineered coupling conditions for perfect transfer are also applied to network setups \cite{Christandl2004PRL92187902PerfectStateTransferQuantum,Christandl2005PRA7132312PerfectTransferArbitraryStates}.
As we show here, a CLS can be perfectly transferred under free evolution after a suitable local phase flip in its amplitude or in selected inter-site hoppings.
The second approach uses optimal control theory \cite{Brif2010NJP1275008ControlQuantumPhenomenaPast,Glaser2015EPJD69279TrainingSchrodingersCatQuantum,Murphy2010PRA8222318CommunicationQuantumSpeedLimit,Zhang2016AoP375435OptimalControlFastHighfidelity,Nakao2017JPBAMOP5065501OptimalControlPerfectState}, where the system is dynamically driven to the target state.
For the tight-binding systems treated here, it aims at maximizing the fidelity of the transfer of CLS across the system by designing smooth time-dependent modulations of the couplings.
The main advantage is its applicability in cases where instantaneous changes, like the phase flips above, are not feasible in practice.
A special representative of optimal control is, e.g., the celebrated stimulated adiabatic Raman passage \cite{Vitanov2017RMP8915006StimulatedRamanAdiabaticPassage} (STIRAP) in three level systems.
Recently \cite{Taie2017ACPSpatialAdiabaticPassageMassive}, CLSs in Lieb lattices have been used, in the form of ``dark states'' \cite{Scully1997QuantumOptics}, as a transfer channel between two local states during a spatial STIRAP process.
We here take an orthogonal viewpoint, since our aim is to transfer the dark state (CLS) itself to other dark states through the network.

\begin{figure}[t] %[t]
\centering
\includegraphics[max size={.9\columnwidth}{1\textheight}]{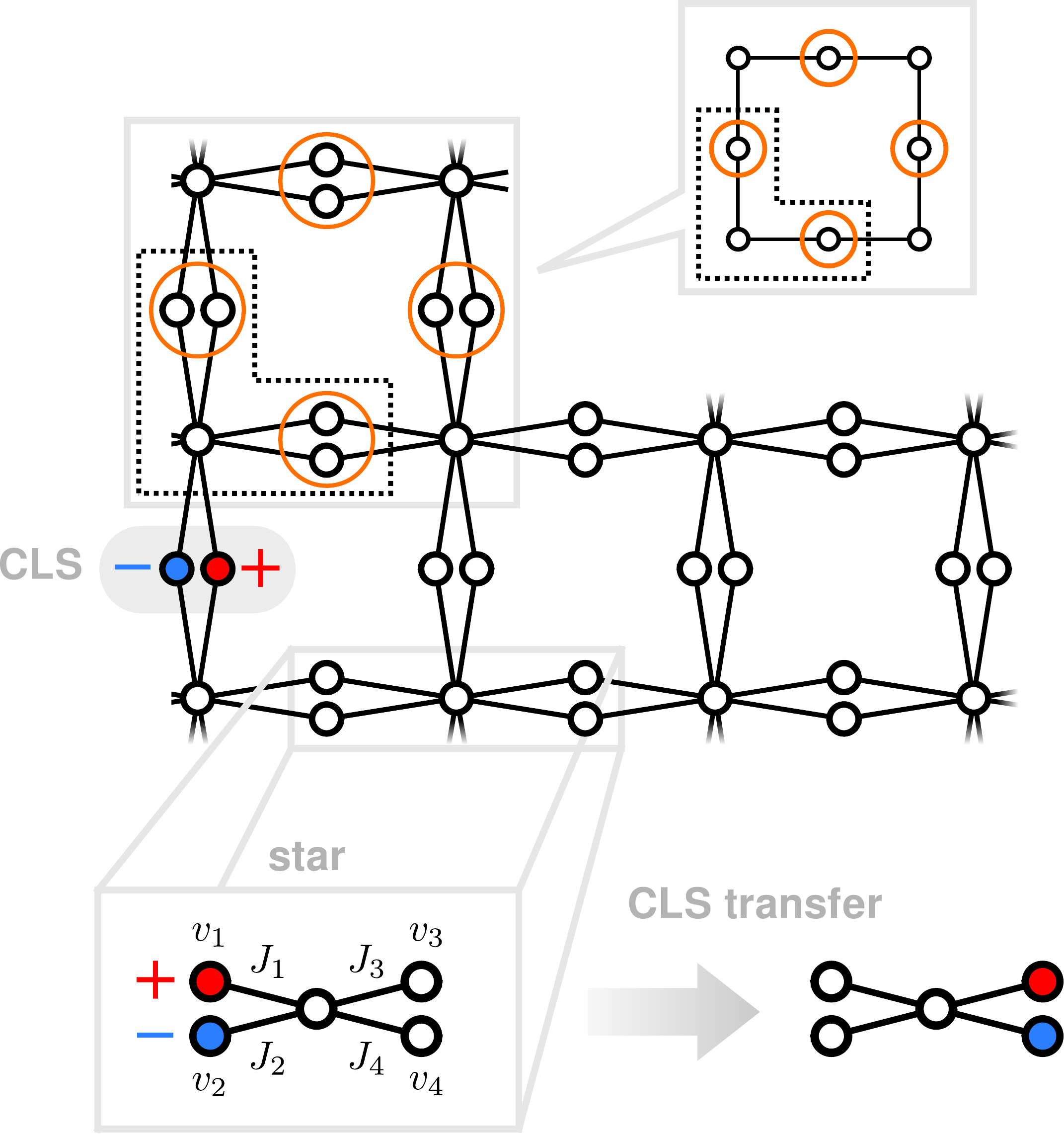}
\caption{
Two-dimensional decorated Lieb lattice (DLL), constructed from the original Lieb lattice (whose plaquette is shown in the upper inset) by replacing the encircled sites with dimers. 
The resulting lattice features two dimers per unit cell (dotted box), and each such dimer can host one CLS with opposite amplitudes on the two dimer sites.
The lower inset shows the isolated ``star'' subsystem functioning as a unit for the CLS transfer.
}
\label{fig:decoratedLieb}
\end{figure}

For definiteness, we shall apply the proposed concept of storage and transfer of CLSs to a decorated Lieb lattice (DLL); see \cref{fig:decoratedLieb}.
It is derived from the original Lieb lattice \cite{Lieb1989PRL621201TwoTheoremsHubbardModel} by replacing the sites of one sublattice with dimers.
The resulting network can be extended to more complex geometries and to higher dimensions, and different CLSs can be routed independently across the network.

%%%%%%%%%%%%%%%%%%%%%%%%%%%%%%%%%%%%%%%%%%%%%%%%%%%%%%%%%%%%%%%%%%%%%%%
%%%%%%%%%%%%%%%%%%%%%%%%%%%%%%%%%%%%%%%%%%%%%%%%%%%%%%%%%%%%%%%%%%%%%%%

\dashSec{Star graph subsystem}The basic building block for CLS transfer in the DLL is its isolated unit cell. It represents a five-point star graph and is shown in the lower inset of \cref{fig:decoratedLieb}.
It is governed by the Hamiltonian
\begin{equation} \label{eq:starHamiltonian}
H = \begin{pmatrix}
 v_1 & 0 & J_{1} & 0 & 0 \\
0 & v_2 & J_{2} & 0 & 0 \\
J_{1} & J_{2} & v_c & J_{3} & J_{4} \\
0 & 0 & J_{3} & v_3 & 0 \\
0 & 0 & J_{4} & 0 & v_4 \\ 
\end{pmatrix}
\end{equation}
with on-site potentials $v_n$ and outer nodes $n=1,2,3,4$ coupled to the central node $c$ by real hoppings $J_{n,c} = J_{c,n} \equiv J_n$.
The above Hamiltonian can be physically realized in various contexts.
One possibility is a coupled waveguide array \cite{Garanovich2012PR5181LightPropagationLocalizationModulated,Szameit2012DiscreteOpticsFemtosecondLaser}, with each node representing a waveguide cross-section and neighboring waveguides evanescently coupled through the overlap of their fundamental modes.
The system is then effectively described by a discrete Schrödinger equation in terms of single-site excitations $\ket{n}$, with time $t$ replaced by the coordinate along the waveguide axis \cite{Szameit2012DiscreteOpticsFemtosecondLaser}.
Another possible physical realization of $H$ is in terms of spins, with each node representing a spin-$\frac{1}{2}$ qubit (measured up or down). 
The Heisenberg XYZ interaction Hamiltonian reduces to this simple description within the Hilbert subspace where total spin is 1 \cite{Bose2007CP4813QuantumCommunicationSpinChain}. 
This subspace is spanned by the state vectors $|k\rangle=|0...010...0\rangle$ where only one spin at position $k$ is up (1) and all others down (0).
Without loss of generality it is possible to consider also any superposition of $|k\rangle$ with $|0...0\rangle$ with all spins down. In this case, the CLS is a two qubit state.

In the presence of local symmetry under permutation only of sites $1$ and $2$, that is, $J_{1} = J_{2} \equiv J$ and $v_{1} = v_{2} \equiv v$, the star Hamiltonian $H$ hosts the eigenstate $\ket{I} = \frac{\ket{1} - \ket{2}}{\sqrt{2}}$.
This is a CLS with opposite amplitudes on sites $1$ and $2$ and vanishing amplitudes on all other sites.
It is thus decoupled by local symmetry from the rest of the star system.
For example, its hopping rate $J_{1} I_{1} + J_{2} I_{2} = J_{1} - J_{2}$ (with $I_{n} = \braket{n|I}$) to the central site vanishes.
It is therefore ``stored'' for an arbitrary time interval until the local symmetry condition is violated.
Crucially, this local-symmetry induced decoupling persists even if $J$ and $v$ are time-dependent, allowing for a gradual modulation $J(t)$ or $v(t)$ without perturbing the CLS [up to a global phase $\ket{I(t)} = e^{i\phi(t) t/\hbar} \ket{I}$ given by $\phi(t) = \int_{t_{0}}^{t} v(t') dt'$].
Moreover, $\ket{I}$ is unaffected by any change of the remainder of the Hamiltonian, i.e, of $v_{c},v_{3},v_{4},J_{3}$ or $J_{4}$.
In complete similarity, for $J_{3} = J_{4}$ and $v_{3} = v_{4}$ the star hosts a second compact localized eigenstate $\ket{F} = \frac{\ket{3} - \ket{4}}{\sqrt{2}}$.
Both $\ket{I}$ and $\ket{F}$ are also dark states since they are the only eigenstates of the Hamiltonian \cref{eq:starHamiltonian} which have non zero coefficients only in two sites.
An important fact is that $\ket{I}$ and $\ket{F}$ are also eigenstates of the full Hamiltonian when the star subsystem is repeatedly connected to form the DLL (see \cref{fig:decoratedLieb}).
In fact, \textit{different} locally symmetric star subsystems, each with its own parameters $v_{n}, J_{n}$, can be connected in a suitable way to form a network hosting multiple independent CLSs, each stored on only two sites with opposite amplitude.

In the following, we first present different protocols transferring a CLS in the star-subsytem.
These can easily be extended to the full DLL, as we will show afterwards.
For the state transfer within the time $T$, the two CLSs $\ket{I}$ and $\ket{F}$ will serve as the initial and final state. Throughout the rest of this work, we set $v_{n} = v$, but $J_{n}$ are not necessarily equal to each other and may also be time-dependent during the pulse.
However, we impose the symmetry condition $J_{3} = J_{4}$ at the end of the transfer to ensure that $\ket{F}$ is an energy eigenstate.
The initially stored CLS is thus transfered to the target location and can be stored again indefinitely.
In a spin network setting, the initial state can be realized as $\ket{I} = \frac{\ket{10\mathbf{0}} - \ket{01\mathbf{0}}}{\sqrt{2}}$, which is a \textit{maximally entangled state} between the spins at sites 1 and 2 of a dimer (while the others summarized by $\mathbf{0}$ are decoupled).
In this context, transfer of state $\ket{I}$ to the corresponding state $\ket{F} = \frac{\ket{\mathbf{0}10} - \ket{\mathbf{0}01}}{\sqrt{2}}$
constitutes transfer of maximal entanglement.

%%%%%%%%%%%%%%%%%%%%%%%%%%%%%%%%%%%%%%%%%%%%%%%%%%%%%%%%%%%%%%%%%%%%%%%
\begin{figure}[t] %[t]
\centering
\includegraphics[max size={.99\columnwidth}{1\textheight}]{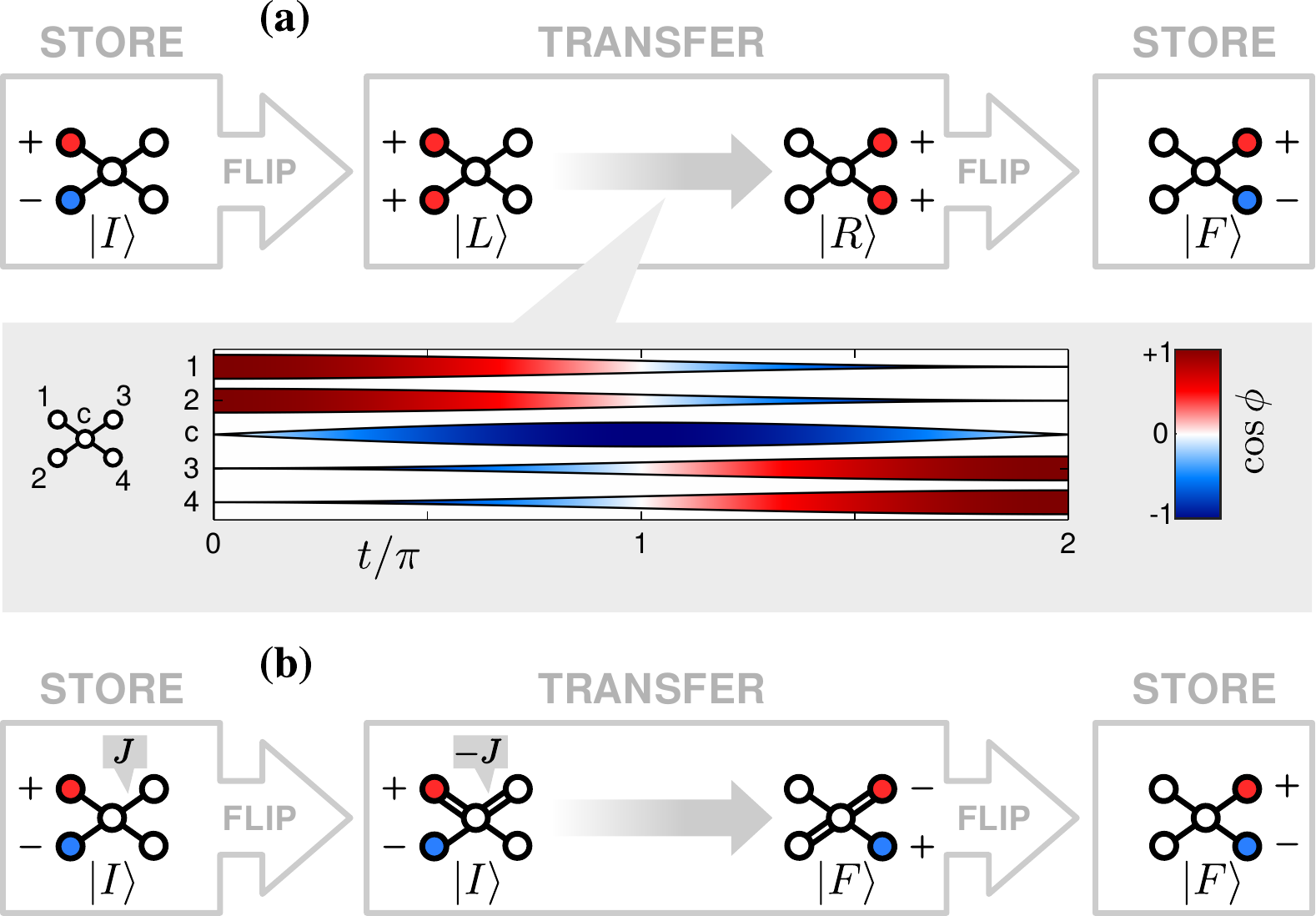}
\caption{
Transfer of a stored dimer-localized CLS $\ket{I}$ to a final CLS $\ket{F}$ within a star subsystem via free time-evolution between sign flips of \textbf{(a)} one component of the initial and final states and \textbf{(b)} one of the couplings of the central site to each of the initial and final CLS dimers.
The inset in (a) shows the evolution of the state $\ket{\psi}$ over time $T = 2\pi$ with $\ket{\psi(0)} = \ket{L}$ and $\ket{\psi(T)} = \ket{R}$ for $v_n = 2J_n = 1/4\, \forall n$.
The phase of $\psi_{i}(t) = \braket{i|\psi(t)}$ at site $i$ is color coded and its amplitude is indicated by the width of the corresponding stripe.
The evolution for (b) is identical except for opposite signs on $\psi_2(t),\psi_{c}(t)$ and $\psi_4(t)$.
}
\label{fig:flipProtocol}
\end{figure}
%%%%%%%%%%%%%%%%%%%%%%%%%%%%%%%%%%%%%%%%%%%%%%%%%%%%%%%%%%%%%%%%%%%%%%%

\dashSec{Transfer by phase flips}For the transfer protocol visualized in \cref{fig:flipProtocol} (a), we set $J_{n} = J$ and consider the possibility to instantaneously imprint a phase flip by $\pi$ (that is, a sign change) on one of the components of $\ket{I}$ at $t = 0$, turning it into $\ket{L} \equiv \frac{\ket{1} + \ket{2} }{\sqrt{2}}$.
This new state $\ket{L}$ is no longer an eigenstate of $H$, and will evolve freely and with unit fidelity within time $T$ to the state $\ket{R} \equiv \frac{\ket{3} + \ket{4}}{\sqrt{2}}$ for suitable chosen on-site potentials and couplings.
Exactly at $t = T$ another sign flip is applied to one of the components of $\ket{R}$ in order to turn it into the desired target CLS $\ket{F}$.
For the choice $J = 1/4$ energy units and $v = 2J$, the transfer time is $T = \frac{\hbar \pi}{2J} =2\pi$ (setting $\hbar = 1$).
General analytical derivations for the evolution $\ket{\psi(t)} = e^{-iHt}\ket{I}$ are given in \cref{appendix:calculations}, exploiting the so-called `equitable partition theorem' \cite{BarrettEquitabledecompositionsgraphs2017,FrancisExtensionsapplicationsequitable2017,Rontgen2018PRB9735161CompactLocalizedStatesFlat}.
As an alternative version of this transfer protocol, we can apply the instantaneous sign flips at $t = 0$ and $T$ to the hoppings $J_1,J_3$ (or $J_2,J_4$) instead, as depicted in \cref{fig:flipProtocol} (b).
The free time evolution is then essentially equivalent to the previous one in \cref{fig:flipProtocol} (a).

\begin{figure}[t] %[t]
\centering
\includegraphics[max size={.99\columnwidth}{1\textheight}]{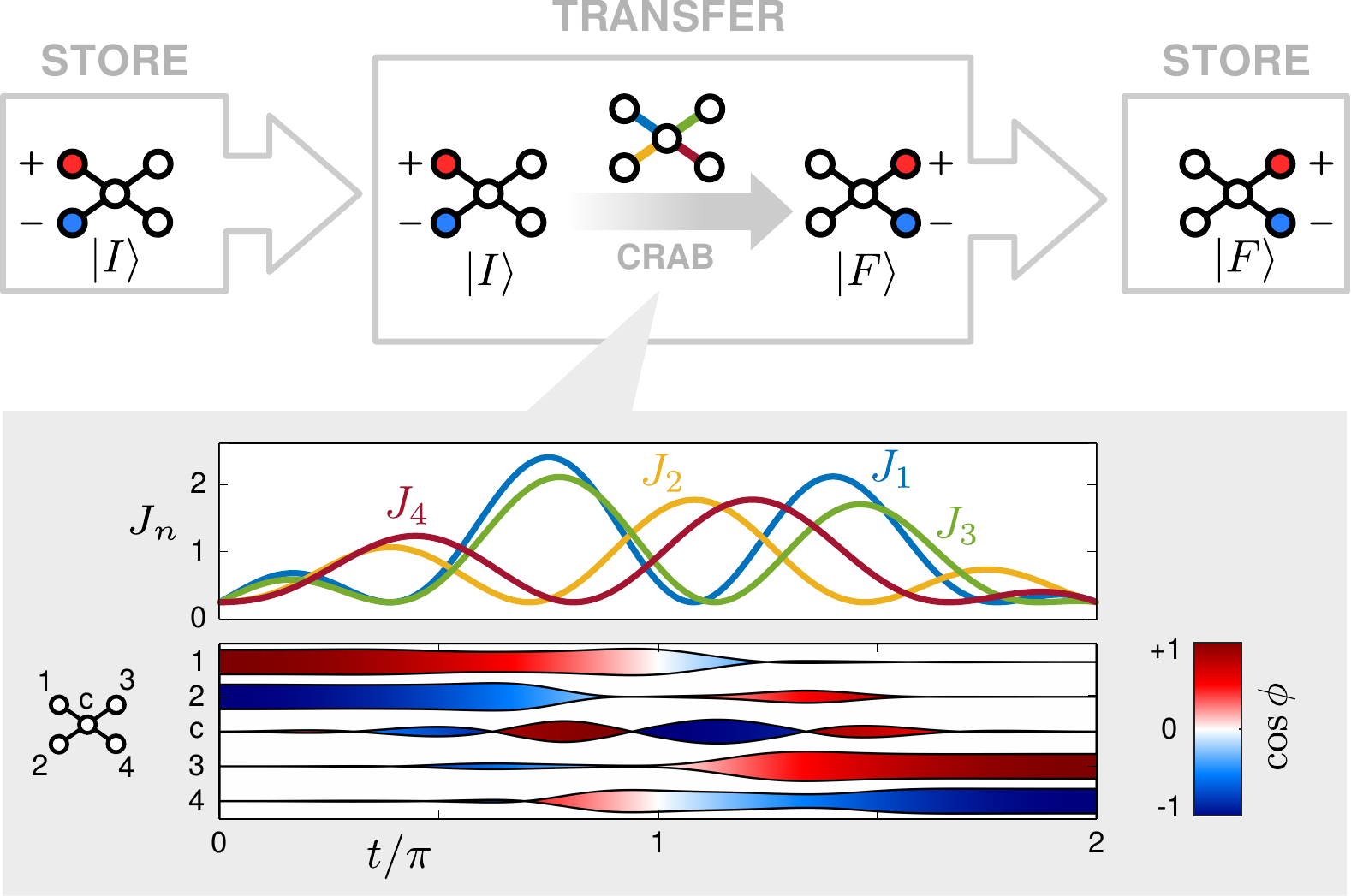}
\caption{
CLS transfer within the star of \cref{fig:flipProtocol} via optimal control using the CRAB method for the couplings $J_n$ of the form in \cref{eq:crabJ}.
At the end of the procedure, all couplings $J_{n} = J$ again.
The inset shows the temporal profile of the $J_n(t)$ and the evolution of the state over time $T=2\pi$.
}
\label{fig:crabProtocol}
\end{figure}

\dashSec{Transfer by optimal control}Now we turn to optimal control solutions in order to design smooth pulses to avoid instantaneous operations.
Taking into account that the initial state $\ket{I}$ is an eigenstate of $H(t=0)$, we need to smoothly drive it out of stationarity in order to end up with the final state $\ket{F}$ as an eigenstate of the final Hamiltonian $H(t=T)$. 
To find smooth optimal driving pulses for the couplings $J_n$ we apply the chopped random-basis\cite{Doria2011PRL106190501OptimalControlTechniqueManyBodya,Caneva2011PRA8422326ChoppedRandombasisQuantumOptimizationa} (CRAB) optimal control method to the functional form
\begin{equation} \label{eq:crabJ}
 J_{n}(t) = J \left\{ 1+\sin\frac{t}{2}\left[x_n\sin(\omega_n t)+x_n'\cos(\omega_n t)\right]^2 \right\} .
\end{equation}
This determines the optimal parameters $x_n,x_n',\omega_n$ ($n=1,2,3,4$) for transferring $\ket{I}$ to $\ket{F}$ in time $T$, with the same initial and final conditions as previously (that is, $J_n = J$).
More information on the CRAB procedure and the specific optimizations performed here can be found in \cref{appendix:calculations}.
For this particular case, we further impose $J = 1/4$ as a minimum value for all $J_n$.
Sign changes in the $J_n$ are thus avoided, which can be advantageous in practice depending on the model's realization (e.g., in the case of waveguide arrays).
The resulting optimal $J_n$-driving pulses are presented in \cref{fig:crabProtocol} together with the state evolution. The infidelity $1 - |\braket{F|\Psi(T_{f})}|^2$ of these pulses is approximately $10^{-10}$.

%%%%%%%%%%%%%%%%%%%%%%%%%%%%%%%%%%%%%%%%%%%%%%%%%%%%%%%%%%%%%%%%%%%%%%%
%%%%%%%%%%%%%%%%%%%%%%%%%%%%%%%%%%%%%%%%%%%%%%%%%%%%%%%%%%%%%%%%%%%%%%%

\dashSec{CLS generation}In the above transfer protocols, we assumed that the initial CLS $\ket{I}$ already exists at $t=0$.
Depending on the system realization, it may indeed be feasible to imprint the CLS directly onto the dimer.
In waveguide arrays, e.g., the two sites could be excited with light of phase difference $\pi$ at the input.
Under some circumstances, however, it is easier to imprint a local excitation only on \textit{one} lattice site rather than to excite the dimer CLS directly.
In particular for spin lattices, the CLS is a (maximally) entangled state, as mentioned above, which may be challenging to imprint directly. 
Therefore, we now propose protocols that start with state $\ket{c}$ (occupying only the central site in the star) at $t=0$ and end with $\ket{I}$ at a desired generation time $T_g$.
To this aim, we initially decouple all outer sites from $\ket{c}$ by setting $J_{n}(t=0)=0$ and turn on only $J_1$ and $J_2$ to a common value $J'$. The on-site potentials are $v_{n} = 2 J = 1/2$ as above.

\begin{figure}[t] %[t]
	\centering
	\includegraphics[max size={.99\columnwidth}{1\textheight}]{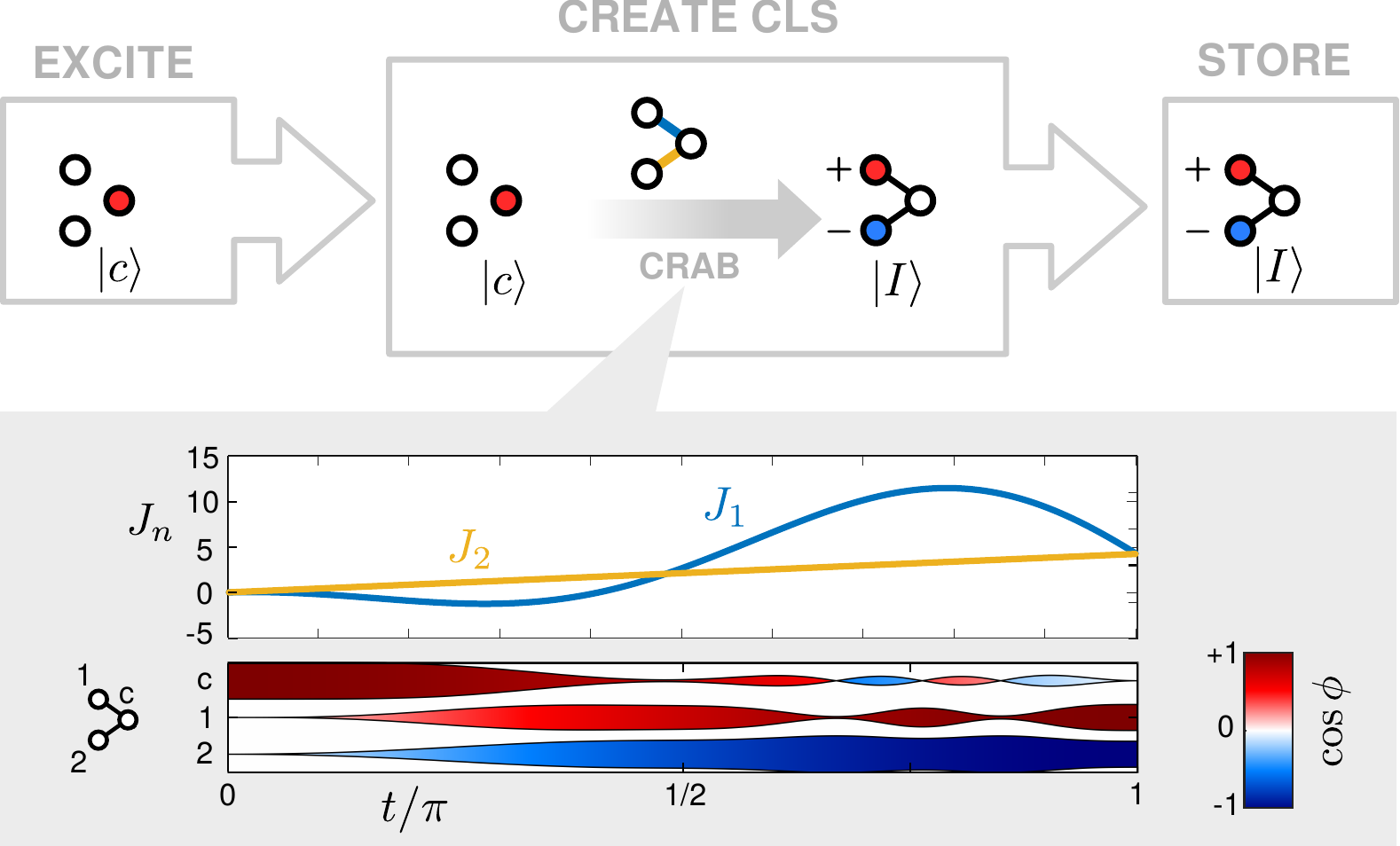}
	\caption{
		Creation of the CLS by an initial excitation of the central site $c$ and subsequent optimal control using the CRAB method for the couplings $J_1$ and $J_{2}$.
		The inset shows the temporal profile of the $J_n(t)$ and the evolution of the state over time $T_{g}=\pi$.
	}
	\label{fig:crabProtocolCLSCreation}
\end{figure}

Analogously to the transfer protocols above, the CLS generation can be done either by instantaneous phase or hopping flips, or by continuous modulation via optimal control.
In the first case, $J_{1}=J_{2}$ are switched on at $t=0$ to the value $J'=3\sqrt{2}$, for which the system evolves freely from $\ket{c}$ to $\ket{L}$ in time $T_g = T/2 = \pi$ (analytical details given in the SM).
Then, at $t=T_g$ a phase flip is applied to obtain $\ket{I}$.
In the second case we use optimal control to evolve the system from $\ket{c}$ directly to $\ket{I}$ by gradually turning on $J_1,J_2$ (from $0$ to $J'$) according to a temporal profile produced via the CRAB method (with an infidelity of approximately $10^{-10}$), graphically depicted in \cref{fig:crabProtocolCLSCreation}.
Here, a linear rise has been chosen for $J_2$, while $J_1$ oscillates and necessarily also becomes negative, as shown in \cref{appendix:calculations}.
With these conditions we prepare and store (due to the equal final couplings) the CLS $\ket{I}$.

As easily anticipated, these CLS generation protocols can also be used in order to perform the state transfer from $\ket{I}$ to $\ket{F}$, now in two steps.
In the first step $\ket{I}$ is evolved to $\ket{c}$ within time $T/2$ using one of the generation protocols \textit{time-reversed}. 
In the second step $\ket{F}$ is generated from $\ket{c}$ within time $T/2$, just like $\ket{I}$ was above.

% In \cref{fig:flipProtocol}(d) we schematically show this protocol as a piecewise transfer scheeme 
% Of course one can again use this protocol for piecewise preparation in the same sence as above schematically shown in \cref{fig:flipProtocol}(f) . 

%%%%%%%%%%%%%%%%%%%%%%%%%%%%%%%%%%%%%%%%%%%%%%%%%%%%%%%%%%%%%%%%%%%%%%%
%%%%%%%%%%%%%%%%%%%%%%%%%%%%%%%%%%%%%%%%%%%%%%%%%%%%%%%%%%%%%%%%%%%%%%%

%%%%%%%%%%%%%%%%%%%%%%%%%%%%%%%%%%%%%%%%%%%%%%%%%%%%%%%%%%%%%%%%%%%%%%%
\begin{figure}[] %[t]
\centering
\includegraphics[max size={1\columnwidth}{.6\textheight}]{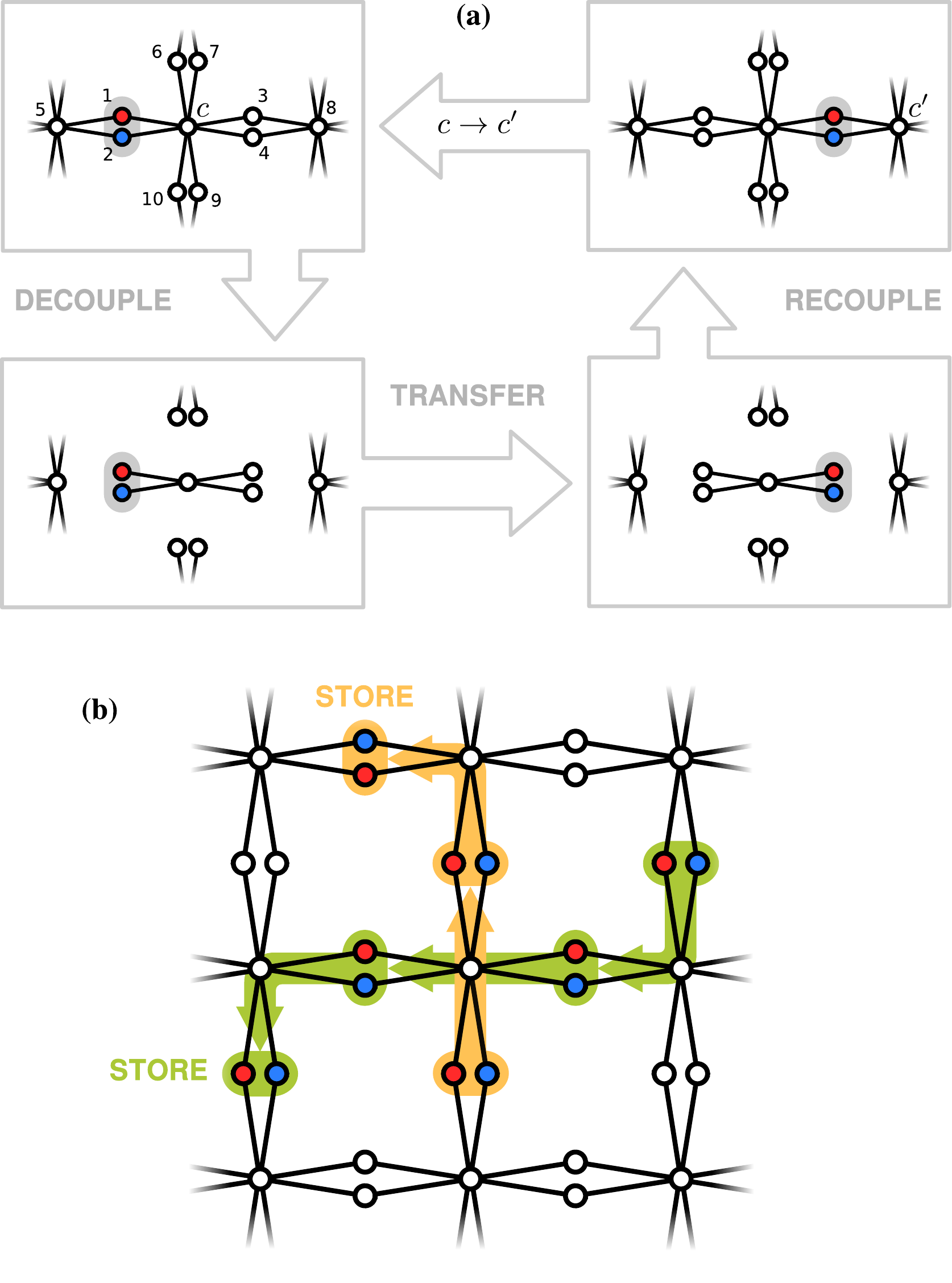}
\caption{
\textbf{(a)} The ``dimer-jump'' process allowing to reduce state transfer within the decorated Lieb-lattice to that within a star subsystem. The cyclic process starts at the upper left and transfers a CLS on sites $1,2$ to another CLS on sites $3,4$ by decoupling the corresponding star from the remainder of the system, performing the actual transfer and subsequently recoupling the star. The process can then start anew, indicated by $c\to c'$.
\textbf{(b)} The robustness of CLSs with respect to perturbations outside their localization domain allows for the simultaneous routing of multiple CLSs along different routes in the network (orange and green directed paths).
As long as the paths of different CLS do not \emph{simultaneously} use the same star-subsystem, there are no restrictions on them.
}
\label{fig:networkTransfer}
\end{figure}
%%%%%%%%%%%%%%%%%%%%%%%%%%%%%%%%%%%%%%%%%%%%%%%%%%%%%%%%%%%%%%%%%%%%%%%

\dashSec{Transfer across a network}
By exploiting the robustness of CLS, the state transfer schemes established above can now be used in the full DLL. 
The elementary process to transfer a compact localized eigenstate within the network is shown in \cref{fig:networkTransfer} (a).
Starting from the upper left, it consists in (i) separating the star subsystem hosting the initial CLS $\ket{I}$ from the remainder of the lattice, (ii) perform the transfer to the final CLS $\ket{F}$ within the star, and (iii) reconnect the star to the remainder. The process can then start anew to transport the state over longer distances.

% We assume that the initial and final state $\ket{I},\ket{F}$ are both CLS, localized on sites $1,2$ and $3,4$, respectively. 
To separate the star while keeping $\ket{I}$ unaffected, we ramp down its outer couplings to surrounding sites to zero within time $\delta t$ such that the local symmetry protecting $\ket{I}$ is \textit{preserved}.
With the site indexing in \cref{fig:networkTransfer} (a), starting at $t=0$ this means that $J_{1,5} (t) = J_{2,5} (t)$ for $0 \leqslant t \leqslant \delta t$ with $J_{1,5}(\delta t)=0$.
This modulation does not perturb $\ket{I}$, even in the limit $\delta t \rightarrow 0$.
The other couplings of the star to its surroundings ($J_{3,8}$, $J_{4,8}$ and $J_{n,c}$ with $n = 6,7,9,10$) can be ramped down in an \textit{arbitrary} way, since they do not connect to the localization domain of $\ket{I}$.
In particular, this can be done even before $J_{1,5}$ and $J_{2,5}$ are ramped down.
Within the separated star, the actual state transfer can be performed according to one of the above protocols over time $T$.
Afterwards, we ramp up the outer couplings again, now preserving the local symmetry which protects $\ket{F}$; 
that is, increasing $J_{3,8}$ and $J_{4,8}$ symmetrically, and the other couplings arbitrarily, from zero to their final values.

Transfer of a CLS to a distant dimer (not in the same star subsystem) is achieved via consecutive ``dimer-jumps'' like the one just described, as depicted in \cref{fig:networkTransfer}\,(b).
This procedure can be employed simultaneously along different paths in the network. These may also intersect in space, as long as in each instant in time the different paths use different star-subsystems; see orange and green paths in \cref{fig:networkTransfer}\,(b).
Once a CLS has reached its final destination, it is stored for arbitrary time. 
In this sense, the proposed network becomes a simple model for a hybrid setup functioning simultaneously as a directional transfer device and as a multiple quantum memory unit.

In practice, an actual realization of an equivalent network will depend on the limitations of underlying physical platform.
We emphasize that the DLL operated on here is a very basic lattice geometry enabling the proposed CLS transfer concept, but can be generalized to other geometries in a straightforward way.
For instance, if intra-dimer coupling is non-negligible due to the small spatial separation (which may be the case in coupled photonic waveguides), an alternative CLS transfer unit can be used, with essentially the same procedure applied, as shown in \cref{appendix:modifiedNetwork}. This transfer unit consists of seven instead of five sites, which allows for a bigger spatial separation between waveguides.

\dashSec{Conclusion}We have demonstrated how local symmetries in a decorated Lieb lattice (DLL) can be exploited to generate, store and transfer compact localized states (CLSs) via different, easily realizable, modulation protocols.
This provides a powerful prototype for a quantum device which simultaneously performs flexible transfer and robust storage of information within the same physical platform.
The transfer protocols utilize either instantaneous phase flips (with unit fidelity) or optimal temporal control of inter-site couplings with near-unit fidelity. 
They can thus be adapted to the needs of different potential realizations in, e.g., electronics, atom-optics, photonics, or acoustics.
An exciting prospect is the application of the concept to decoherence-free transfer and storage of maximally entangled spin states, represented here by CLSs.
Under the very weak requirement of local symmetries protecting the CLSs, extension of the proposed concept to alternative network geometries and different dimensionality is straightforward.
Based on multiple intersecting CLS transfer paths as proposed here, a future vision would be the design of a dynamical network with switchable quantum gates and embedded quantum memories.
\begin{acknowledgments}
	Financial support by the Deutsche Forschungsgemeinschaft under grant DFG Schm 885/29-1 is gratefully acknowledged. 
	M.R. gratefully acknowledges financial support by the `Stiftung der deutschen Wirtschaft' in the framework of a scholarship.
	I.B gratefully acknowledges financial support by IKY (Greek State Scholarship Foundation).
\end{acknowledgments}

\appendix
\widetext
\section{State transfer and CLS-creation protocols presented in the main part} \label{appendix:calculations}
\subsection{State transfer (phase flip)}
For $v_{n} = v$, $n \in \{1,2,3,4,c\}$ and $J_{n} = J$, $n \in \{1,2,3,4\}$, \cref{eq:starHamiltonian} can be analyzed analytically by means of the so-called equitable partition theorem \cite{Rontgen2018PRB9735161CompactLocalizedStatesFlat,BarrettEquitabledecompositionsgraphs2017,FrancisExtensionsapplicationsequitable2017}.
This theorem gives a block-partitioning of a locally symmetric Hamiltonian, allowing for a simpler computation of both eigenvectors and eigenvalues. The underlying local symmetry is restricted to be a permutation that commutes with the Hamiltonian and which acts non-trivially only on a subset of sites.
For the Hamiltonian treated here, the underlying symmetry is described by the operator $S$ performing the cyclic permutation $1 \to 2 \to 3 \to 4 \to 1$ and mapping $c$ to itself. Then, $[H,S] = 0$, and by the equitable partition theorem, the eigenvectors are
\begin{equation*}
\ket{\phi^{(i)}} = \left\{ \begin{pmatrix}
-1/\sqrt{2}\\1/\sqrt{2}\\0\\0\\0
\end{pmatrix},
\begin{pmatrix}
0\\0\\0\\-1/\sqrt{2}\\1/\sqrt{2}
\end{pmatrix},
\begin{pmatrix}
1/2\\1/2\\0\\-1/2\\-1/2
\end{pmatrix},
\begin{pmatrix}
1/(2\sqrt{2})\\1/(2\sqrt{2})\\-1/\sqrt{2}\\1/(2\sqrt{2})\\1/(2\sqrt{2})
\end{pmatrix},
\begin{pmatrix}
1/(2\sqrt{2})\\1/(2\sqrt{2})\\1/\sqrt{2}\\1/(2\sqrt{2})\\1/(2\sqrt{2})
\end{pmatrix} \right\}
\end{equation*}
with corresponding eigenvalues $E^{(i)} = \{ v,v,v,v-2J,v+2J \}$.
In terms of these eigenvectors, the two states $\ket{L} = (1/\sqrt{2},1/\sqrt{2},0,0,0)^{T}$ and $\ket{R} = (0,0,0,1/\sqrt{2},1/\sqrt{2})^{T}$ are expanded as
\begin{align*}
\ket{L} &= \sum_{i} a^{(i)}_{l} \ket{\phi^{(i)}} = \frac{1}{2} \ket{\phi^{(3)}} + \frac{1}{2} \ket{\phi^{(4)}} + \frac{1}{2} \ket{\phi^{(5)}} \\
\ket{R} &= \sum_{i} a^{(i)}_{r} \ket{\phi^{(i)}} = -\frac{1}{2} \ket{\phi^{(3)}} + \frac{1}{2} \ket{\phi^{(4)}} + \frac{1}{2} \ket{\phi^{(5)}}
\end{align*}
with coefficients $a^{(i)}_{l,r}  = \braket{\phi^{(i)}|l,r}$.
To achieve a unitary time-evolution of state $\ket{L}$ at $t = 0$ to $\ket{R}$ at $t = T_{f}$, the conditions $a^{(i)}_{l} e^{-i E^{(i)} T_{f} /\hbar}= a^{(i)}_{r}$ must hold, leading to
\begin{equation} \label{eq:timeEvolutionSystem}
e^{-i E^{(3)} T_{f} /\hbar} = -1,\; e^{-i E^{(3)} T_{f} /\hbar} = 1,\; e^{-i E^{(5)} T_{f}/\hbar} = 1
\end{equation}
which is fulfilled for
\begin{equation} \label{eq:solutionsProtocol1}
v = J \left( \frac{4 k_{1}}{1+2k_{2}} -2\right),\; T_{f} = \frac{\pi \hbar (1 + 2k_{2})}{2J},\; k_{1,2} \in \mathbb{Z}.
\end{equation}
For $k_{1} =1,\; k_{2} = 0$, one finds $v = 2 J, T_{f} = \frac{\pi \hbar}{2 J}$ which was given in the main part of this work.

\subsection{State transfer (hopping flip)}
For $v_{n}$ as above and $J_{1} = J_{3} = J$ and $J_{2} = J_{4} = -J$, the eigenvectors of $H$ are
\begin{equation*}
\ket{\phi^{(i)}} = \left\{ \begin{pmatrix}
1/\sqrt{2}\\1/\sqrt{2}\\0\\0\\0
\end{pmatrix},
\begin{pmatrix}
0\\0\\0\\1/\sqrt{2}\\1/\sqrt{2}
\end{pmatrix},
\begin{pmatrix}
1/2\\-1/2\\0\\-1/2\\1/2
\end{pmatrix},
\begin{pmatrix}
1/(2\sqrt{2})\\-1/(2\sqrt{2})\\-1/\sqrt{2}\\1/(2\sqrt{2})\\-1/(2\sqrt{2})
\end{pmatrix},
\begin{pmatrix}
1/(2\sqrt{2})\\-1/(2\sqrt{2})\\1/\sqrt{2}\\1/(2\sqrt{2})\\-1/(2\sqrt{2})
\end{pmatrix} \right\}
\end{equation*}
with corresponding eigenvalues $E^{(i)} = \{ v,v,v,v-2J,v+2J \}$.
In order to enable a unitary time-evolution of state $\ket{L} = (1/\sqrt{2},-1/\sqrt{2},0,0,0)^{T}$ at $t = 0$ to $\ket{R} = (0,0,0,1/\sqrt{2},-1/\sqrt{2})^{T}$ at $t = T_{f}$, one must solve $a^{(i)}_{l} e^{-i E^{(i)} T_{f} /\hbar}= a^{(i)}_{r}$.
This system of equations is equal to \cref{eq:timeEvolutionSystem}, and the parameters needed to achieve perfect transfer are thus identical to that of the phase flip protocol.

\subsection{Preparation-storage and piecewise transfer (phase flip)}
Let us now show how one can yield $\ket{L}$ from initially exciting the state $\ket{c} = (0,0,1,0,0)^{T}$ for $J_{3}= J_{4} = 0$ and $J_{1} = J_{2} = J'$. In this case, the Hamiltonian \cref{eq:starHamiltonian} is block-diagonal, and the eigenvectors split into two subspaces. The eigenvectors relevant for the expansion of $\ket{L}$ and $\ket{c}$ are
\begin{equation*}
\ket{\phi^{(i)}} = \left\{
\begin{pmatrix}
1/\sqrt{2}\\-1/\sqrt{2}\\0\\0\\0
\end{pmatrix},
\begin{pmatrix}
-1/2\\-1/2\\1/\sqrt{2}\\0\\0
\end{pmatrix},
\begin{pmatrix}
1/2\\1/2\\1/\sqrt{2}\\0\\0
\end{pmatrix}
\right\}
\end{equation*}
with eigenvalues $E^{(i)}= \{v,v-\sqrt{2}J,v + \sqrt{2}J\}$.
By expanding $\ket{c}$ and $\ket{L}$ into the eigenvectors of the Hamiltonian, one can show that for perfect state transfer
\begin{equation*}
e^{-i E^{(2)} T_{f} /\hbar} = -1,\; e^{-i E^{(3)} T_{f} /\hbar} = 1
\end{equation*}
must be fulfilled. This is the case for
\begin{equation}
v = \frac{\sqrt{2}J' (4k'_{1} -1)}{1 + 4k'_{2}},\, T_{f} = \frac{\pi \hbar (4k'_{1} -1)}{2v}\;\;\; \text{or} \;\;\; v = \frac{\sqrt{2}J' (4k'_{1} + 1)}{-1 + 4k'_{2}},\, T_{f} = \frac{\pi \hbar (4k'_{1} + 1)}{2v}
\end{equation}
where in both equations $k'_{1,2} \in \mathbb{Z}$. The special form $v = \frac{\sqrt{2} J'}{3}, t = \frac{\pi \hbar}{2 v}$ given in the main part of this work is obtained by taking the second solution as well as setting $k'_{1} = 0,k'_{2} = 1$ and $J' = 3\sqrt{2}J$.

\subsection{Preparation-storage and piecewise transfer (hopping flip)}
In this case, we have $J_{2} = -J = - J_{1}$. The eigenvectors relevant for the expansion of $\ket{L}$ and $\ket{c}$ are thus
\begin{equation*}
\ket{\phi^{(i)}} = \left\{
\begin{pmatrix}
1/\sqrt{2}\\1/\sqrt{2}\\0\\0\\0
\end{pmatrix},
\begin{pmatrix}
-1/2\\1/2\\1/\sqrt{2}\\0\\0
\end{pmatrix},
\begin{pmatrix}
1/2\\-1/2\\1/\sqrt{2}\\0\\0
\end{pmatrix}
\right\}
\end{equation*}
with eigenvalues $v,v-\sqrt{2}J,v + \sqrt{2}J$. The expansion of $\ket{c}$ and $\ket{L}$ is equal to that of the phase flip protocol, with identical parameters needed to achieve perfect transfer.

\subsection{Methods and calculations for optimal control protocols}

The optimal control method Chopped Random-Basis quantum optimization CRAB \cite{Doria2011PRL106190501OptimalControlTechniqueManyBodya,Caneva2011PRA8422326ChoppedRandombasisQuantumOptimizationa} is based on expressing the time-dependent driving functions/pulses for the control fields (here the couplings J) on a truncated randomized basis.
This recasts the problem of a functional minimization (of the infidelity function) to a multi-variable function
minimization that can be performed, for example, via a direct-search method. Here we use Nelder-Mead optimization for the parameters of the function.
To achieve state transfer within the star-subsystem, we assume an initial and final $J_0=1/4$ which is also the minimum threshold for all couplings, and we have chosen the following (CRAB-inspired) expressions for the driving functions:
\begin{equation*} 
J_{n}(t) = J_{0} \left\{ 1+\sin\left(\frac{t}{2}\right)\left[x_n\sin(\omega_n t)+x_n'\cos(\omega_n t)\right]^2 \right\}
\end{equation*}
The method essentially starts with a random set of frequencies $w_i$ ($i=1,...,4$) which is different in every iteration and  optimizes $8$ amplitude parameters $x_i,x'_{i}$. The set $\{w_i,x_i,x'^{i}\}$  which minimizes the infidelity is defining the optimal pulses.
For the pulses presented in \cref{fig:crabProtocol} we have $x_{n} = \{0.5850, 2.4015, 2.5033, 0.2199\}$, $x'_{n} = \{2.9997,0.5954, 0.4555, 2.8103 \}$ and $\omega_{n} = \{1.4452, 1.3069, 1.1680, 1.3510 \}$.

For the preparation storage and piecewise transfer protocol with optimal control, shown in \cref{fig:crabProtocolCLSCreation}, we use 
\begin{align*}
J_{1} &= \Big\{ 1+x \sin(\omega t)+x' \sin(\omega' t) \Big\} 3 \sqrt{2} (1 - t/\pi ) \\
J_{2} &= 3 \sqrt{2} (1 - t/\pi ) .
\end{align*}
Here we have a set of 4 parameters to optimize (both frequencies and amplitudes) and we allow the $J_{n}$ to take also negative values.
For \cref{fig:crabProtocolCLSCreation}, the optimal parameter values are $x = 0.8292$, $x' = 1.5246$, $\omega = 1.7638$, $\omega' = 1.9434$.
The necessity of negative values for $J_{n}$ is apparent if we see the system of differential equations which governs the dynamics which simplifies to:
\begin{align*}
\dot{\psi}_1=i(J_{1}\psi_c+v_1 \psi_1)\\
\dot{\psi}_2=i(J_{2}\psi_c+v_2 \psi_2)\\
\dot{\psi}_c=i(J_{1}\psi_1+J_{2}\psi_2+v_c \psi_c)
\end{align*}
With $\ket{\psi(t=0)} = \ket{c}$ and $v_1=v_2=v_c=1/2$, the amplitude $\psi_1$ or $\psi_2$ can acquire the desired negative value $-1/\sqrt{2}$ only if $J_1$ or $J_2$ becomes negative (for some $t$-intervals) as well, respectively.

\section{Modified network for evanescently coupled waveguide arrays} \label{appendix:modifiedNetwork}

%%%%%%%%%%%%%%%%%%%%%%%%%%%%%%%%%%%%%%%%%%%%%%%%%%%%%%%%%%%%%%%%%%%%%%%
\begin{figure}[ht!] %[t]
	\centering
	\includegraphics[max size={0.5\columnwidth}{.6\textheight}]{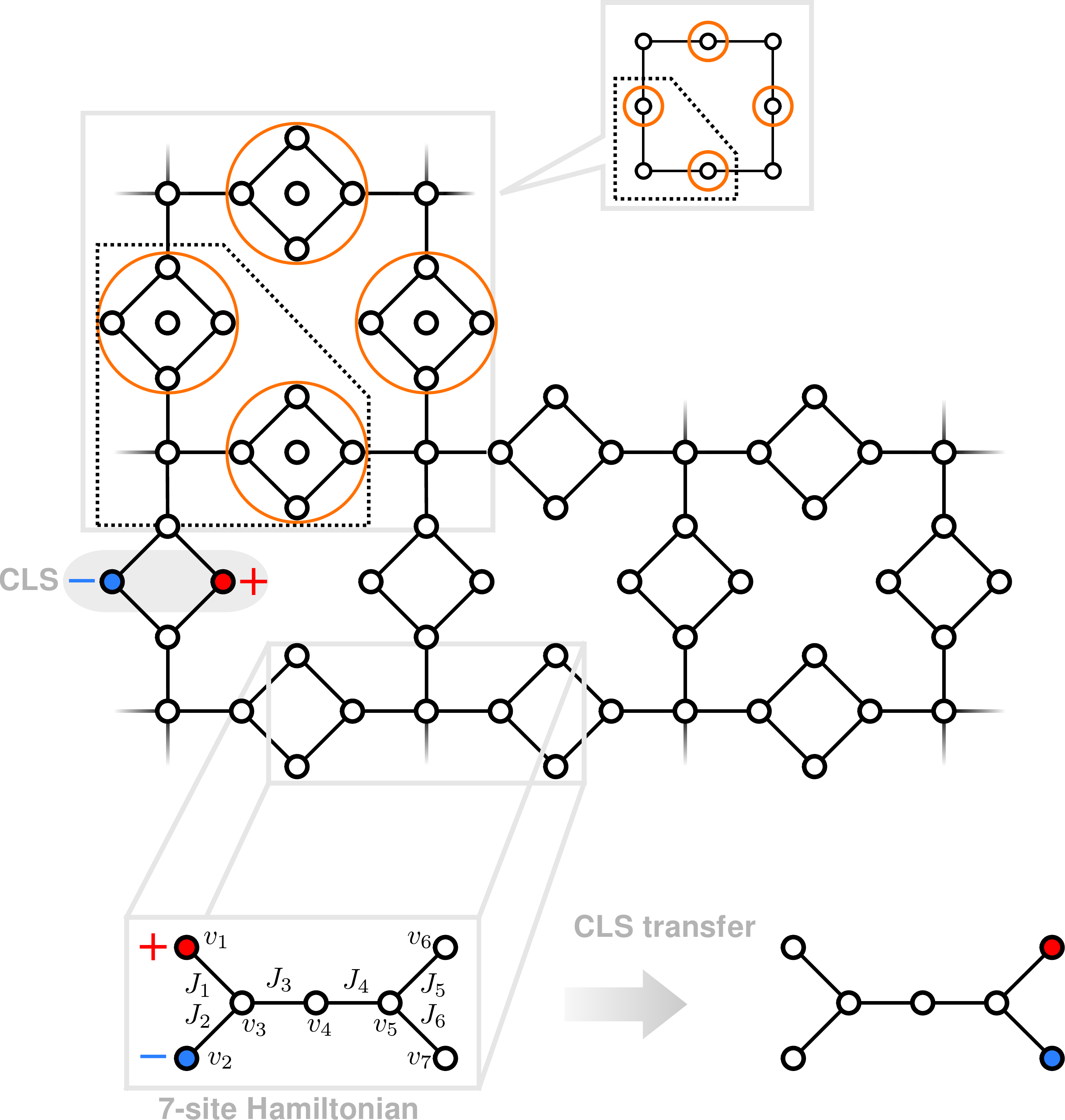}
	\caption{
		Modified network for evanescently coupled waveguide arrays.
	}
	\label{fig:modifiedLiebLattice}
\end{figure}
%%%%%%%%%%%%%%%%%%%%%%%%%%%%%%%%%%%%%%%%%%%%%%%%%%%%%%%%%%%%%%%%%%%%%%%
We here provide a modified network, along with some transfer protocols, for cases where the five-point Hamiltonian \cref{eq:starHamiltonian} is not suitable for a physical realization. The basis for these protocols is the seven-point graph, described by the Hamiltonian
\begin{equation} \label{eq:sevenPointGraph}
H_{7}=\begin{pmatrix}
v_1 & 0 & J_{1} & 0 & 0 & 0 & 0 \\
0 & v_2 & J_{2} & 0 & 0 & 0 & 0 \\
J_{1} & J_{2} & v_{3} & J_{3} & 0 & 0 & 0 \\
0 & 0 & J_{3} & v_{4} & J_{4} & 0 & 0 \\
0 & 0 & 0 & J_{4} & v_{5} & J_{5} & J_{6} \\
0 & 0 & 0 & 0 & J_{5} & v_6 & 0 \\
0 & 0 & 0 & 0 & J_{6} & 0 & v_7 \\ 
\end{pmatrix}.
\end{equation}
For $v_{1} = v_{2}, J_{1} = J_{2}$ and $v_{3} = v_{4}, J_{3} = J_{4}$, this Hamiltonian hosts two compact localized states, $\ket{I} = (1/\sqrt{2},-1/\sqrt{2},0,0,0,0,0)^{T}$ and $\ket{F} = (0,0,0,0,0,1/\sqrt{2},-1/\sqrt{2})^{T}$. We assume that at $t = 0$, the state $\ket{I}$ is excited and one wants to transfer it to the CLS $\ket{F}$ at $t = T_{f}$.

\subsubsection{Phase flip protocol}
Similar to protocol 1 for the five-point Hamiltonian, we assume that at $t = 0$, a phase flip is applied, changing the state $\ket{I}$ into $\ket{L} = (1/\sqrt{2},1/\sqrt{2},0,0,0,0,0)^{T}$. Our aim is to transfer this into $\ket{R} = (0,0,0,0,0,1/\sqrt{2},1/\sqrt{2})^{T}$ at $t = T_{f}$, where we apply another phase flip, turning $\ket{R}$ into $\ket{F}$.
In the following, we will choose $v_{i} = 0, \; i = \{1,\ldots{},7\}$, $J_{1} = J_{2} = J_{5} = J_{6} = J$.
With this choice of parameters, $H_{7}$ is globally symmetric w.r.t. to a left-right flip with site $4$ as a center for $J_{3} = J_{4}$, but only locally symmetric for $J_{3} \ne J_{4}$. In both cases, the so-called `nonequitable partition theorem' \cite{Rontgen2018PRB9735161CompactLocalizedStatesFlat,Fritscher2016SJMAA37260ExploringSymmetriesDecomposeMatrices} allows to obtain analytical expressions for both the eigenvalues and eigenvectors of $H_{7}$.
This theorem is similar to the equitable partition theorem, but allows for the treatment of a greater class of local symmetries. For the current example, the eigenvalues of $H_{7}$ are the union of the eigenvalues of
\begin{equation*}
R = \begin{pmatrix}
v & \sqrt{\xi} J_{3} & 0 & 0\\
\sqrt{\xi} J_{3} & v & J & J\\
0 & J & v & 0\\
0 & J & 0 & v
\end{pmatrix},\;\;
C_{0} = \begin{pmatrix}
v & J & J\\
J & v & 0\\
J & 0 & v
\end{pmatrix}
\end{equation*}
with $\xi = J_{3}^2 + J_{4}^2$.
If we denote the eigenvectors $\textbf{x}^{\nu}, \; \nu \in \{1,\ldots{},4\}$ of $R$ as $(x_{1}^{\nu},\ldots{},x_{4}^{\nu})^T$ and those of $C_{0}$ as $(w_{1}^{\mu},w_{2}^{\mu},w_{3}^{\mu})^{T},\; \mu \in \{1,2,3\}$, with $T$ denoting the transpose, then the (unnormalized) eigenvectors of $H_{7}$ are
\begin{equation}
\begin{pmatrix}
\frac{J_{3}}{\sqrt{\xi}} x_{4}^{\nu}\\
\frac{J_{3}}{\sqrt{\xi}} x_{3}^{\nu}\\
\frac{J_{3}}{\sqrt{\xi}} x_{2}^{\nu}\\
x_{1}^{\nu}\\
\frac{J_{4}}{\sqrt{\xi}} x_{2}^{\nu}\\
\frac{J_{4}}{\sqrt{\xi}} x_{3}^{\nu}\\
\frac{J_{4}}{\sqrt{\xi}} x_{4}^{\nu}
\end{pmatrix},\;\;
\begin{pmatrix}
w_{3}^{\mu}\\
w_{2}^{\mu}\\
w_{1}^{\mu}\\
0\\
-\frac{J_{3}}{J_{4}} w_{1}^{\mu}\\
-\frac{J_{3}}{J_{4}} w_{2}^{\mu}\\
-\frac{J_{3}}{J_{4}} w_{3}^{\mu}
\end{pmatrix}.
\end{equation}
For $J_{3} = J_{4} = J' = \sqrt{3}$, $J= 1$, we find that
\begin{equation*}
\ket{\phi^{(i)}} = \left\{ \begin{pmatrix}
1/\sqrt{2}\\-1/\sqrt{2}\\0\\0\\0\\0\\0
\end{pmatrix},
\begin{pmatrix}
0\\0\\0\\0\\0\\1/\sqrt{2}\\-1/\sqrt{2}
\end{pmatrix},
\begin{pmatrix}
-\sqrt{3}/4\\-\sqrt{3}/4\\0\\1/2\\0\\-\sqrt{3}/4\\-\sqrt{3}/4
\end{pmatrix},
\begin{pmatrix}
-\eifrac{1}{2 \sqrt{2}}\\\ -\eifrac{1}{2\sqrt{2}} \\1/2\\0\\-1/2\\\eifrac{1}{2\sqrt{2}} \\ \eifrac{1}{2\sqrt{2}}
\end{pmatrix},
\begin{pmatrix}
-\eifrac{1}{2 \sqrt{2}}\\-\eifrac{1}{2 \sqrt{2}}\\-1/2\\0\\1/2\\\eifrac{1}{2 \sqrt{2}}\\\eifrac{1}{2 \sqrt{2}}
\end{pmatrix},
\begin{pmatrix}
\eifrac{1}{4 \sqrt{2}}\\\eifrac{1}{4 \sqrt{2}}\\-1/2\\\sqrt{3/8}\\-1/2\\\eifrac{1}{4 \sqrt{2}}\\\eifrac{1}{4 \sqrt{2}}
\end{pmatrix},
\begin{pmatrix}
\eifrac{1}{4 \sqrt{2}}\\\eifrac{1}{4 \sqrt{2}}\\1/2\\ \sqrt{3/8}\\1/2\\\eifrac{1}{4 \sqrt{2}}\\\eifrac{1}{4 \sqrt{2}}
\end{pmatrix}
\right\}
\end{equation*}
are normalized eigenvectors of \cref{eq:sevenPointGraph} with corresponding eigenvalues $E^{(i)} = \{0,0,0,-\sqrt{2},\sqrt{2},-2\sqrt{2},2\sqrt{2}\}$. In terms of $\ket{\phi^{(i)}}$, the initial and final state $\ket{l,r}$ are written as
\begin{align*}
\ket{L} &= \sum_{i} a^{(i)}_{l} \ket{\phi^{(i)}} =  - \sqrt{\frac{3}{8}} \ket{\phi^{(3)}} - \frac{1}{2} \ket{\phi^{(4)}} - \frac{1}{2} \ket{\phi^{(5)}} + \frac{1}{4} \ket{\phi^{(6)}} + \frac{1}{4} \ket{\phi^{(7)}} \\
\ket{R} &= \sum_{i} a^{(i)}_{r} \ket{\phi^{(i)}} = - \sqrt{\frac{3}{8}} \ket{\phi^{(3)}} + \frac{1}{2} \ket{\phi^{(4)}} + \frac{1}{2} \ket{\phi^{(5)}} + \frac{1}{4} \ket{\phi^{(6)}} + \frac{1}{4} \ket{\phi^{(7)}}
\end{align*}
where $a^{(i)}_{l,r} =  \braket{\phi^{(i)}|l,r}$.
Then, for $T_{f} = \frac{\pi \hbar}{\sqrt{2}}$, the state $\ket{L}$ has evolved into $\ket{R}$. We then apply another instantaneous phase flip, turning $\ket{R}$ into the CLS $\ket{F}$.

\subsubsection{Hopping flip protocol}
This protocol is similar to the hopping flip protocol presented in the main part of this work for the five-point Hamiltonian. We initially set $v_{i} = 0, \;i = \{1,\ldots{},7\}$, $J_{1} = J_{2,} = J_{5} = J_{6} = J$ and $J_{3} = J_{4} = J'$ as for protocol 1b.
Then, at $t = 0$, the couplings $J_{2}$ and $J_{6}$ are instantaneously changed to $-J$. After this change, the new eigenvectors of \cref{eq:sevenPointGraph} are
\begin{equation*}
\ket{\phi^{(i)}} = \left\{ \begin{pmatrix}
1/\sqrt{2}\\1/\sqrt{2}\\0\\0\\0\\0\\0
\end{pmatrix},
\begin{pmatrix}
0\\0\\0\\0\\0\\1/\sqrt{2}\\1/\sqrt{2}
\end{pmatrix},
\begin{pmatrix}
-\sqrt{3}/4\\\sqrt{3}/4\\0\\1/2\\0\\-\sqrt{3}/4\\\sqrt{3}/4
\end{pmatrix},
\begin{pmatrix}
\eifrac{1}{2 \sqrt{2}}\\-\eifrac{1}{2 \sqrt{2}}\\-1/2\\0\\1/2\\-\eifrac{1}{2 \sqrt{2}}\\\eifrac{1}{2 \sqrt{2}}
\end{pmatrix},
\begin{pmatrix}
\eifrac{1}{2 \sqrt{2}}\\-\eifrac{1}{2 \sqrt{2}}\\1/2\\0\\-1/2\\-\eifrac{1}{2 \sqrt{2}}\\\eifrac{1}{2 \sqrt{2}}
\end{pmatrix},
\begin{pmatrix}
-\eifrac{1}{4 \sqrt{2}}\\\eifrac{1}{4 \sqrt{2}}\\1/2\\-\sqrt{3/8}\\1/2\\-\eifrac{1}{4 \sqrt{2}}\\\eifrac{1}{4 \sqrt{2}}
\end{pmatrix},
\begin{pmatrix}
-\eifrac{1}{4 \sqrt{2}}\\\eifrac{1}{4 \sqrt{2}}\\-1/2\\ -\sqrt{3/8}\\-1/2\\-\eifrac{1}{4 \sqrt{2}}\\\eifrac{1}{4 \sqrt{2}}
\end{pmatrix}
\right\}
\end{equation*}
with corresponding eigenvalues $E^{(i)} = \{ 0,0,0,-\sqrt{2},\sqrt{2},-2\sqrt{2},2\sqrt{2} \}$.
In terms of these eigenvectors, the initial and final states $\ket{I,F}$ are expanded as
\begin{align*}
\ket{I} &= \sum_{i} a^{(i)}_{I} \ket{\phi^{(i)}} =   - \sqrt{\frac{3}{8}} \ket{\phi^{(3)}} + \frac{1}{2} \ket{\phi^{(4)}} + \frac{1}{2} \ket{\phi^{(5)}} - \frac{1}{4} \ket{\phi^{(6)}} - \frac{1}{4} \ket{\phi^{(7)}} \\
\ket{F} &= \sum_{i} a^{(i)}_{F} \ket{\phi^{(i)}} = - \sqrt{\frac{3}{8}} \ket{\phi^{(3)}} - \frac{1}{2} \ket{\phi^{(4)}} - \frac{1}{2} \ket{\phi^{(5)}} - \frac{1}{4} \ket{\phi^{(6)}} - \frac{1}{4} \ket{\phi^{(7)}} 
\end{align*}
where $a^{(i)}_{I,F} =  \braket{\phi^{(i)}|I,F}$.
Then, for $T_{f} = \frac{\pi \hbar}{\sqrt{2}}$, the state $\ket{I}$ has evolved into $\ket{F}$. We then perform another flip such that the couplings $J_{2} = J_{6} = J$, and $\ket{F}$ becomes a CLS again.

\subsubsection{State transfer with CRAB}
\begin{figure}[ht] %[t]
	\centering
	\includegraphics[max size={.5\columnwidth}{1\textheight}]{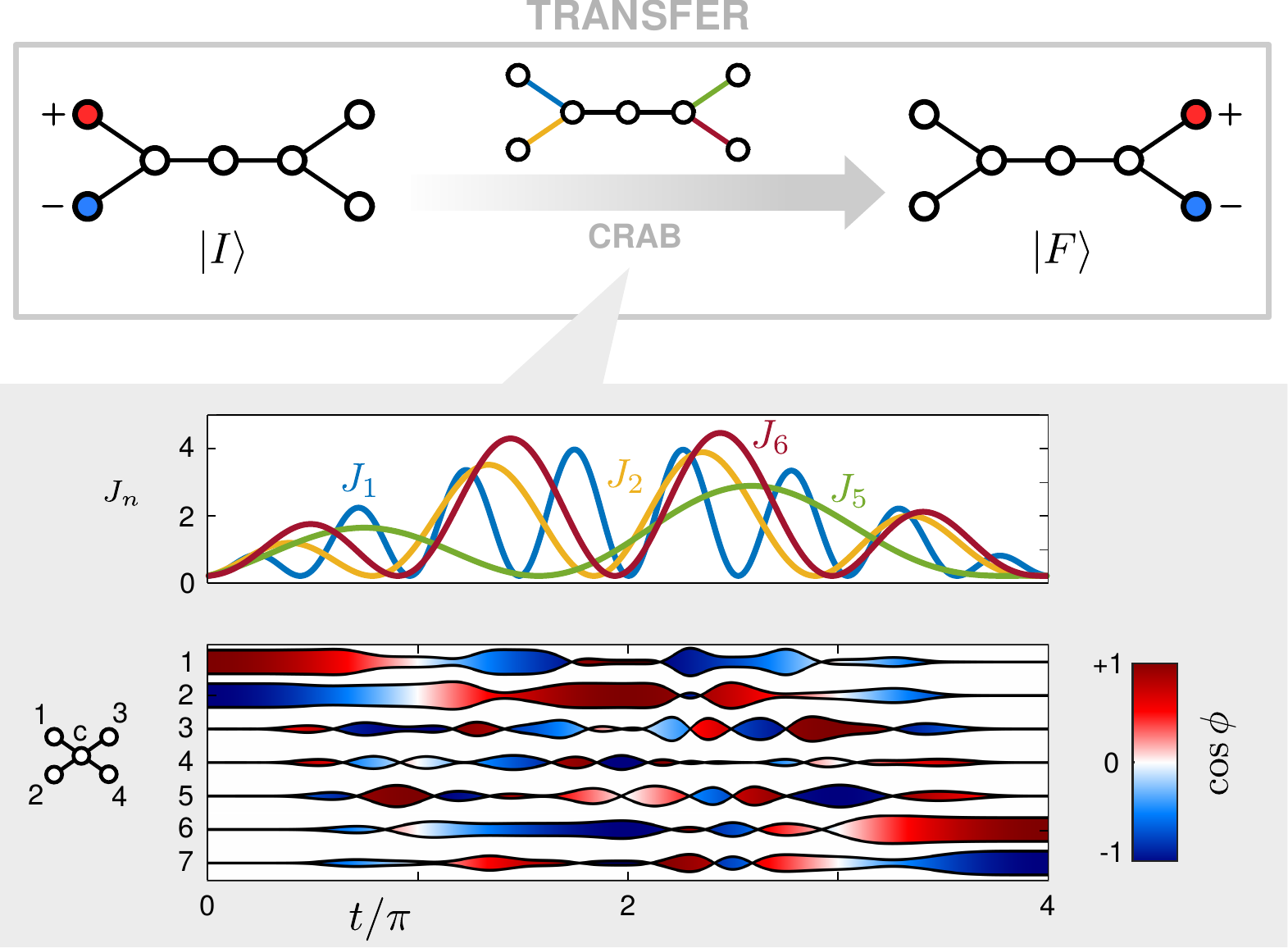}
	\caption{
		CLS transfer within the seven-point graph of \cref{fig:modifiedLiebLattice} via optimal control using the CRAB method for the couplings $J_1$, $J_{2}$, $J_{5}$ and $J_{6}$ of the form in \cref{eq:sevenSiteCRABJ}. $J_{3} = J_{4} = 3$ are constant throughout the process.
		The inset shows the temporal profile of the varied $J_n(t)$ and the evolution of the state over time $T=4\pi$.
	}
	\label{fig:sevensiteCLSTransfer}
\end{figure}

This protocol is similar to the five-site optimal control protocol presented in the main part of this work. It is graphically depicted in \cref{fig:sevensiteCLSTransfer}.
We start with an initial CLS $\ket{I}$, with Hamiltonian parameters $v_{i} = 1/2$ and $J_{1} = J_{2} = J_{5} = J_{6} = \frac{1}{4 \sqrt{2}} = J$ as well as $J_{3} = J_{4} = 3$. We then vary $J_{n}, n = \{1,2,5,6\}$ in time according to
\begin{equation} \label{eq:sevenSiteCRABJ}
J_{n} = J \Big\{ 1+\sin \frac{t}{4} \Big[ x_{n} \sin(\omega_{n} t)+x_{n}' \cos(\omega_{n} t)\Big] ^2 \Big\}.
\end{equation}
The optimal parameters are $x_{n} = \{ 4.1435, 3.2435, 2.5509, 4.7169\}$, $x'_{n} = \{2.2124, 3.3942, 3.3221, 1.9491 \}$ and $\omega_{n} = \{1.9171, 0.9476, 0.4496, 0.9671\}$, and the infidelity $1 - |\braket{F|\Psi(T_{f})}|^2$ of these pulses is approximately $10^{-8}$.

%%%%%%%%%%%%%%%%%%%%%%%%%%%%%%%%%%%%%%%%%%%%%%%%%%%%%%%%%%%%%%%%%%%%%%%
%%%%%%%%%%%%%%%%%%%%%%%%%%%%%%%%%%%%%%%%%%%%%%%%%%%%%%%%%%%%%%%%%%%%%%%

\subsubsection{Piecewise transfer and CLS creation with CRAB}
\begin{figure}[ht] %[t]
	\centering
	\includegraphics[max size={.5\columnwidth}{1\textheight}]{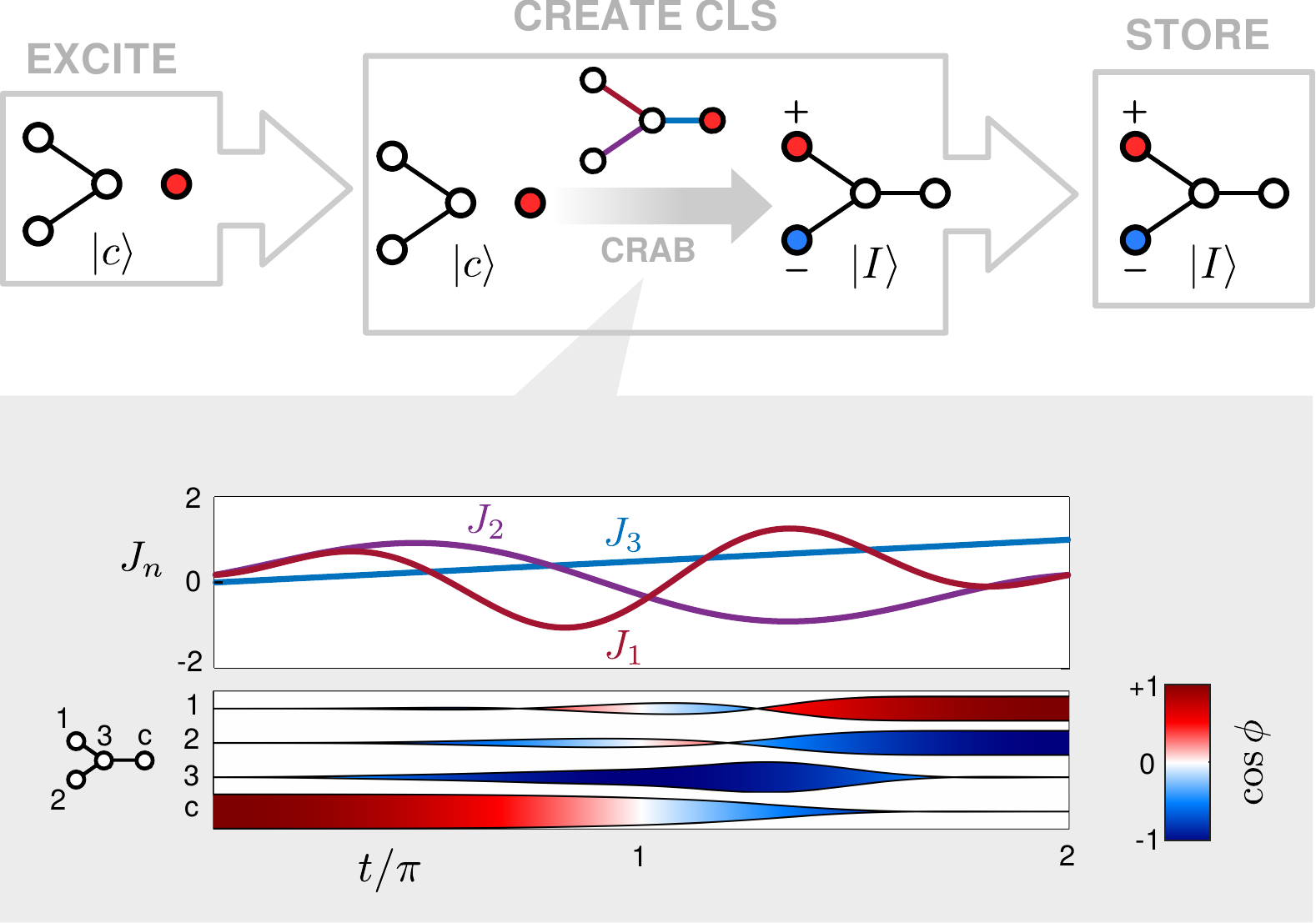}
	\caption{
		Creation of the CLS by an initial excitation of the central site $c$ and subsequent optimal control using the CRAB method for the couplings $J_1, J_{2}$ and $J_{3}$. The inset shows the temporal profile of the $J_n(t)$ and the evolution of the state over time $T_{g}=2\pi$.
	}
	\label{fig:sevensiteCLSCreation}
\end{figure}

This protocol is similar to the three-site optimal control protocol presented in the main part of this work used to prepare a CLS, as was shown in \cref{fig:crabProtocolCLSCreation}. It is graphically depicted in \cref{fig:sevensiteCLSCreation}, with an infidelity of approximately $10^{-8}$.
We start with an excitation of site $4 \equiv c$, which is decoupled from the remainder of the graph. The coupling $J_{3} = t/(2\pi)$ to the left half of the graph is then linearly ramped up, and the couplings $J_{1,2}(t)$ are varied as well. At time $T_{g} = 2 \pi$, the CLS $\ket{I}$ is created. The on-site potentials $v_{i} = 1/2$ are constant throughout the process, and the ansatz for $J_{1,2}(t)$ is
\begin{align} \label{eq:fourSiteCRABJ}
J_{n} &= J \Big\{ 1+\sin \frac{t}{2}  \Big[ (x_{n} \sin(\omega_{n} t)+x'_{n} \cos(\omega_{n} t)) \Big] \Big\},\; n \in \{1,2\}
\end{align}
with $J = \frac{1}{4 \sqrt{2}}$ as above and final parameters $x_{n} = \{6.9763, 4.1098\}$, $x'_{n} =  \{2.1072,6.4490 \} $ and $\omega_{n} = \{1.7465,0.7946 \}$.
Again, as for the protocol shown in the main part, one could use this protocol also to transfer $\ket{I}$ to $\ket{F}$ by first transferring $\ket{I}$ to $\ket{c}$ by means of a time-reversed reversal of the protocol shown, and then transfer $\ket{c}$ to $\ket{F}$ via its forward version.
%%%%%%%%%%%%%%%%%%%%%%%%%%%%%%%%%%%%%%%%%%%%%%%%%%%%%%%%%%%%%%%%%%%%%%%
%%%%%%%%%%%%%%%%%%%%%%%%%%%%%%%%%%%%%%%%%%%%%%%%%%%%%%%%%%%%%%%%%%%%%%%

%

\end{document}